\documentclass[fleqn,usenatbib]{mnras}

\usepackage{newtxtext,newtxmath}

\usepackage[T1]{fontenc}

\DeclareRobustCommand{\VAN}[3]{#2}
\let\VANthebibliography\thebibliography
\def\thebibliography{\DeclareRobustCommand{\VAN}[3]{##3}\VANthebibliography}

\usepackage{graphicx}	
\usepackage{amsmath}	
\usepackage{orcidlink}
\usepackage{xcolor}
\usepackage[percent]{overpic}
\usepackage{tabularx} 
\definecolor{red}{rgb}{0.96, 0.36, 0.36}

\title[MW--LMC Validation - Simulation Based Inference]{A Simulation Based Inference Approach to the Dynamics of the MW--LMC System - Validation}
\author[R. A. N. Brooks et al.]{Richard A. N. Brooks$^{1,2}$\thanks{E-mail: richard.brooks.22@ucl.ac.uk}\orcidlink{0000-0001-5550-2057},
Jason L. Sanders$^{1}$\orcidlink{0000-0003-4593-6788},
Adam M. Dillamore$^{1}$\orcidlink{0000-0003-0807-5261},
Nicolás Garavito-Camargo$^{2,3}$\orcidlink{0000-0001-7107-1744}, \newauthor
Adrian~M.~Price-Whelan$^{2}$\orcidlink{0000-0003-0872-7098}
\\
$^{1}$Department of Physics and Astronomy, University College London, London, WC1E 6BT, UK\\
$^{2}$Center for Computational Astrophysics, Flatiron Institute, Simons Foundation, 162 Fifth Avenue, New York, NY 10010, USA\\
$^{3}$Steward Observatory, University of Arizona, 933 North Cherry Avenue, Tucson, AZ 85721, USA\\
}

\date{Accepted XXX. Received YYY; in original form ZZZ}

\pubyear{\the\year{}}

\begin{document}
\label{firstpage}
\pagerange{\pageref{firstpage}--\pageref{lastpage}}
\maketitle

\begin{abstract}
The infall of the LMC into the Milky Way (MW) has generated dynamical disequilibrium throughout the MW. The interaction has displaced the MW's centre of mass, manifesting as an apparent ‘reflex motion’ in velocities of outer halo stars. Often, expensive high fidelity MW--LMC simulations are required to model these effects, though the range of model parameter spaces can be large and complex. 
We investigate the ability of lower fidelity, rigid MW--LMC simulations to reliably infer the model parameters of higher fidelity $N$-body and hydrodynamical cosmological zoom-in MW--LMC simulations using a Simulation Based Inference (SBI) approach. 
We produce and release a set of $128,000$ MW--LMC rigid potentials, with stellar haloes evolved to present-day, each adopting a unique combination of model parameters including the MW mass, the LMC mass and the dynamical friction strength. For these simulation parameters, we use SBI to find their posterior distributions.
We find that our SBI framework trained on rigid MW--LMC simulations is able to correctly infer the true simulation LMC mass within a $1\sigma$ confidence interval from both $N$-body and cosmological simulations when knowledge of the induced MW reflex motion is provided as data. 
This motivates future applications of the presented SBI framework to observational data, which will help constrain both MW and LMC properties, as well as the dynamics of the MW's reflex motion.
\end{abstract}

\begin{keywords}
Galaxy: kinematics and dynamics -- Galaxy: halo -- Galaxy: evolution -- Magellanic Clouds -- software: machine learning -- software: simulations
\end{keywords}



\section{Introduction}\label{sec:introduction}

The Milky Way (MW) is undergoing a merger with the LMC\footnote{The Large Magellanic Cloud.} \citep[see][for a comprehensive review detailing the effects of the LMC on the MW]{Vasiliev2023}. The LMC is thought to be on its first pericentric passage and to have a dark matter mass $M_{\mathrm{LMC}}\sim 10^{11}\,\mathrm{M}_{\odot}$ \citep{Besla2007, Besla2010, Boylan-Kolchin2011, Penarrubia2016, Kravtsov2024}. An alternative scenario has the LMC on its second pericentric passage, although most features of the earlier passage are superseded by the more recent passage at a smaller pericenter \citep{Vasiliev2024}. 
Many Local Group phenomena require a large mass for the LMC to explain: for example, the kinematics of its globular clusters \citep{Watkins2024}, the kinematics of MW satellites \citep{CorreaMagnus2022, Kravtsov2024}; dynamical models of MW stellar streams \citep{2019MNRAS.487.2685E, Koposov2019, Shipp2021, Vasiliev2021, Koposov2023}; and the timing argument (\citealt{Penarrubia2016}, but see also \citealt{Benisty:2022, Chamberlain:2023, Benisty2024}) all require an LMC mass $M_{\mathrm{LMC}}\sim 10-20\, \times 10^{10}\,\mathrm{M}_{\odot}$ \citep[see fig.~1 of][for a summary of LMC mass estimates]{Vasiliev2023}. 
The orbit of the LMC is sensitive to the assumed Galactic potential \citep[see fig.~3 of][]{Vasiliev2023} and, because the LMC is of considerable mass, it is also subject to dynamical friction from the MW dark matter halo \citep{Chandrasekhar1943}. 

At present-day the LMC is at a Galactocentric distance $d = 49.6 \pm 0.5 \,\mathrm{kpc}$ \citep{Pietrzynski2019},
and heliocentric line of sight velocity of $v_{\mathrm{los},} = 262.2 \pm 3.4\,\mathrm{km}\,\mathrm{s}^{-1}$ \citep{vanderMarel2014}. There is more ambiguity in the measurement of the position of the LMC centre \citep[see sec.~4.1,][]{vanderMarel2014} which correspond to different reported values of the mean proper motions \citep{GaiaCollaboration2018, GaiaCollaboration2021, Wan2020} due to both perspective effects and internal motion in the LMC. Table.~2 in \citet{Vasiliev2023} gives an overview of some of the recent measurements of the mean LMC proper motions and the associated centre. Although, more recently, hyper velocity stars that are ejected via interactions with the LMC's supermassive black hole have been used to constrain both the LMC's dynamical centre and past orbit \citep[][]{Han2025, Lucchini2025}.

As the LMC, a satellite galaxy, falls into the gravitational potential of the MW, the central galaxy, responds by generating a density wake \citep{Chandrasekhar1943}. This is because an infalling satellite will have a broad range of orbital frequencies that resonate with dark matter particles of the host galaxy \citep{Mulder1983, Weinberg1986}. 
The classical `conic' wake trailing the LMC is described as the \textit{transient response}, whereas the response elsewhere in the MW halo is the \textit{collective response} caused by the barycentre displacement \citep{Garavito-Camargo2019, 2021ApJ...919..109G, Tamfal2021, Foote2023}. 
The density wake of the LMC is predicted to leave an observable signature in the density and kinematics of MW halo stars \citep[e.g.,][]{Conroy2021, Cavieres2024, Chandra2025, Yaaqib2024, Bystrom2025, Amarante2024}.

As the LMC is massive, $M_{\mathrm{LMC}}\sim 10-20\, \times 10^{10}\,\mathrm{M}_{\odot}$ (a 1 : 5--10 mass ratio with the MW), and it has just passed its pericentre with a relative
velocity $ > 300 \,\mathrm{km}\,\mathrm{s}^{-1}$ \citep{Hunt2025}, this has caused significant dynamical disequilibrium in the Galaxy. Specifically, the inner and outer parts of the MW halo have experienced different strengths of gravitational acceleration due to the LMC. To the Galactocentric observer, this appears as if the Galactic Northern sky is red-shifted and the Southern sky is blue-shifted as the halo moves preferentially ‘up’, towards the Galactic North. 
This displacement, known as the \textit{‘reflex motion'}, manifests itself as a dipole signal in density \citep{Belokurov2019, Garavito-Camargo2021, Conroy2021, Amarante2024} that is higher in the Galactic North, and similarly there exists a dipole in stellar radial velocities \citep{2019MNRAS.487.2685E, 2020MNRAS.494L..11P, Petersen2021, Erkal2021, Yaaqib2024, Chandra2025b, Chandra2025, Bystrom2025, Sheng2025}.
The magnitude of the velocity dipole is called the travel velocity, $v_{\rm{travel}}$, and its orientation is called the \textit{apex} direction of the reflex motion $(l_{\rm{apex}}$, $b_{\rm{apex}})$ in Galactic coordinates. Although recent studies agree that the travel velocity direction and magnitude should point towards a point on the past orbit of the LMC, they do not converge on their values \citep[e.g.,][fig.~9]{Bystrom2025}. This is likely due to each study using different stellar tracers over varying radial ranges and sky coverages.

The MW--LMC system can be modelled at varying levels of simulation fidelity. There exists a trade off between the simulation fidelity and the ability to explore large model parameter spaces \citep[e.g., see,][in the context of modelling the Sagittarius stream]{Vasiliev2021}.
The simplest prescription are \textit{rigid} models of the MW and LMC galaxies. Rigid models describe the MW and LMC as analytic potentials that have fixed functional forms i.e., they are time-invariant, although they are allowed to move in response to each other. The main advantage of these models is their computational inexpense ($\sim\mathcal{O}(\rm seconds)$). However, the main drawback is their inability to truly capture the physics of galaxy formation e.g., the galaxy masses will change over time, which changes the gravitational potential, and hence their resulting interaction. 
To accommodate many further complexities, the MW--LMC system can instead be modelled using $N$-body simulations in combination with Basis Function Expansions \citep[BFEs, e.g.,][]{Lilley2018a, Lilley2018b, Sanders2020, 2020MNRAS.494L..11P, Garavito-Camargo2019, Garavito-Camargo2021, Lilleengen2023, Vasiliev2024}. These models are often labelled as idealised or \textit{deforming} MW--LMC simulations. They aim to match the present-day conditions of the MW and LMC while also accounting for the deformations to both galaxies due to the re-distribution of dark matter due to the LMC's infall. This includes the formation of a dark matter dynamical friction wake trailing the LMC \citep{Chandrasekhar1943}. These deformations are not something that the simpler rigid MW--LMC models can capture although the dynamical friction can still be accounted for. However, they are significantly more computationally expensive to run ($\sim\mathcal{O}(\rm months)$) as they contain billions of particles. As such, the parameter space explored for MW and LMC properties is much smaller e.g., only a handful of models exist to vary the LMC's mass \citep{Garavito-Camargo2019}.
Further, there have been MW--LMC pairs identified within hydrodynamical cosmological simulations. For example, there are many cases within the \textit{Feedback In Realistic Environments} \citep[FIRE,][]{Wetzel2023} Latte, the \textit{Milky Way-est} \citep{Buch2024}, the \textit{DREAMS} \citep{Rose2025} and the \textit{Auriga} simulation suites \citep{Grand2017, Grand2024}. This is by far the most physically realistic setting to study MW--LMC interactions. However, in a similar vein to deforming simulations, these cosmological simulations can be computationally expensive and cover somewhat limited model parameter spaces. 

Simulation Based Inference \citep[SBI, see,][for a recent review]{Cranmer2020} offers a route to strike a balance between simulation fidelity and the exploration of model parameter space. 
SBI is a powerful statistical framework for performing inference in complex modelling scenarios where traditional analytic methods are impractical or impossible. It is particularly well-suited for estimating parameter posterior distributions when the likelihood function cannot be explicitly defined, or is intractable from simulation.
Instead, forward simulations are used to generate samples of the data and parameters. 
This technique is also known as \textit{‘likelihood free inference}' or \textit{‘distribution free inference'}. Although, these names are slightly misleading as the inference is not likelihood-free, rather we avoid explicitly defining a likelihood and instead model it using many forward simulations. 
For our application to the dynamics of the MW--LMC interaction, removing the requirement to provide a well-defined likelihood is beneficial as we are no longer biased by assuming some functional form e.g., a Gaussian distribution for the MW reflex parameters:~$v_{\rm{travel}}$, $l_{\rm{apex}}$, and $b_{\rm{apex}}$ \citep{Petersen2021, Yaaqib2024, Bystrom2025, Chandra2025}. This is especially important in the context of cosmological MW--LMC analogues as both the model parameter space and the posteriors can be non-trivial. 
SBI has been applied across many fields of research: from mathematics \citep[e.g.,][]{Verdier2022}, to quantitative finance \citep[e.g.,][]{Yu2012}, to seismology \citep[e.g.,][]{Saoulis2024}, therefore owing itself to become an effective \textit{lingua franca} for facilitating communication across scientific domains. Furthermore, SBI has been increasingly used throughout astrophysics problems \citep[e.g.,][]{Weyant2013, Alsing2019, Jeffrey2021, Lemos2021, Hermans2021, vonWietersheim-Kramsta2024, Lovell2024, Widmark2025, Sante2025, XiangyuanMa2025, Jeffrey2025}. 

For the first time, we will demonstrate the validity of using SBI to infer properties of the MW--LMC system when using measured dynamical quantities, i.e., the reflex motion, from many forward models \rm{of} the MW--LMC system using rigid prescriptions. We will investigate the ability of our SBI framework to infer the true parameter values of MW--LMC simulations described by rigid, deforming and cosmological models. This will demonstrate the ability that SBI has to upscale the fidelity of simpler simulations to more complex ones. Ultimately, this allows more a rapid exploration of large model parameters spaces at the fraction of the computational cost required to run the higher fidelity simulations. We will show that our SBI framework is reliable and unbiased through a series of diagnostic tests. Our novel approach will be applied to real data with the aim of constraining properties of the MW, LMC and the dynamics of the induced MW reflex motion. 

The plan of the paper is as follows. 
In Sec.~\ref{sec:simulations}, we give detailed descriptions of the low-fidelity rigid MW--LMC simulations we use throughout this study.
In Sec.~\ref{sec:sbi} we describe the SBI framework, detailing the use of Bayesian statistics and flexible machine learning models for parameter inference. 
In Sec.~\ref{sec:validation}, we validate our SBI framework by confirming the ability to recover known simulation parameters from existing rigid, deforming and cosmologically motivated MW--LMC simulations. 
Further, in Sec.~\ref{sec:posterior-diagnostic-checks}, we perform diagnostic checks on the posterior distributions estimated using SBI.
In Sec.~\ref{sec:discussion}, we discuss our results, any potential caveats and assess the ability to apply SBI to observed MW stellar halo data to infer MW--LMC model parameters.
Finally, we summarise and conclude our results in Sec.~\ref{sec:conclusions}.

\section{Simulations}\label{sec:simulations}

This section provides the details for our low fidelity rigid MW--LMC simulations. Throughout this work, conversions between the heliocentric and Galactocentric cartesian frames use the right-handed Cartesian coordinate system with the Sun positioned at $\vec{r}_{\odot} = (-8.3, 0.0, 0.02)\, \rm {kpc}$ \citep[][]{GRAVITYCollaboration2019, Bennett2019}, with
velocity $\vec{v}_{\odot} = (11.1, 244.24, 7.24)\, \rm {km}\,\rm s^{-1}$ \citep[][]{Schonrich2010, Eilers2019}. 

\subsection{The Milky Way -- LMC potentials}\label{sec:simulations-MWLMCpot}

We use the galaxy dynamics C++/Python package \texttt{agama} \citep{2019MNRAS.482.1525V} to generate many rigid MW--LMC simulations, each with a unique combination of model parameters. We describe the models for the MW and LMC in the following sections.

\subsubsection{The Milky Way}\label{sec:simulations-MW}

Our MW model consists of a spherical bulge, an exponential stellar disc, and a dark matter halo. We keep the parameters of the stellar distributions fixed and close to the commonly used values suggested by \cite{McMillan2017}. The bulge density profile is:

\begin{equation}
\rho_b \propto (1 + r / r_b)^{-\Gamma} \exp \left[ - (r / u_b)^2 \right],
\end{equation}
\noindent
with $r_b = 0.2\,\mathrm{kpc}$, $u_b = 1.8\,\mathrm{kpc}$, $\Gamma = 1.8$, and total mass $1.2 \times 10^{10} \, \mathrm{M}_{\odot}$. The disc follows the exponential density profile:

\begin{equation}
\rho_d \propto \exp \left[ - R / R_d \right] \operatorname{sech}^2 \left[ z / (2h) \right].
\end{equation}
\noindent
with a total mass $5 \times 10^{10} \, \mathrm{M}_{\odot}$, a scale radius $R_d = 3\,\mathrm{kpc}$, and a scale height $h_d = 0.4\,\mathrm{kpc}$.
We use a Navarro-Frenk-White \citep[NFW,][]{Navarro1996, Navarro1997} dark matter halo density profile described by $M_{200,\rm MW}$ and $c_{200}$. These quantities are defined by a sphere enclosing an overdensity that is 200 times the critical density of the Universe, $\rho_{\mathrm{crit}} = 3H^{2}/8\pi G$, as denoted by the ‘$200$' subscript.
We constrain the normalisation of the halo mass profile such that the circular velocity at the solar position is approximately $235\,\mathrm{km}\,\mathrm{s}^{-1}$ \citep[e.g., matching constraints from][within the associated uncertainty]{McMillan2017}. In combination with a chosen $M_{200,\rm MW}$ value, this allows us to find the corresponding concentration parameter, $c_{200}$. 

\subsubsection{The LMC}\label{sec:simulations-lmc}

We model the LMC as a Hernquist dark matter halo \citep{Hernquist1990}. 
We impose the observational constraint from \citet{vanderMarel2014} that the derived rotation curve peaks at $91.7 \pm \, 18.8 \,\mathrm{km}\,\mathrm{s}^{-1}$ at $8.7\,\mathrm{kpc}$, which corresponds to an enclosed dynamical mass of $M(<8.7\,\mathrm{kpc}) = 1.7 \pm 0.7 \times 10^{10}\,\mathrm{M_{\odot}}$. The mass enclosed of a Hernquist profile within a given radius is $\mathrm{M}(<r)= \mathrm{M_{tot}} \,r^2 / (r+a)^2$ \citep{Hernquist1990}. Hence, for a given total and enclosed LMC mass, we can solve for, $a$, the scale radius.

\subsubsection{The MW--LMC interaction}\label{sec:simulations-mwlmcinteraction}

The trajectories of the MW and LMC under their mutual gravitational attraction are found by numerically integrating the equations:

\begin{gather}\label{equ3-6}
    \dot{\mathbf{x}}_{\mathrm{MW}} = \mathbf{v}_{\mathrm{MW}} \\
    \dot{\mathbf{v}}_{\mathrm{MW}} = -\nabla \Phi_{\mathrm{LMC}} (\mathbf{x}_{\mathrm{MW}} - \mathbf{x}_{\mathrm{LMC}}) \\ 
    \dot{\mathbf{x}}_{\mathrm{LMC}} = \mathbf{v}_{\mathrm{LMC}} \\
    \dot{\mathbf{v}}_{\mathrm{LMC}} = -\nabla \Phi_{\mathrm{MW}} (\mathbf{x}_{\mathrm{LMC}} - \mathbf{x}_{\mathrm{MW}}) + \lambda_{\mathrm{DF}} \mathbf{a}_{\mathrm{DF}},
\end{gather}
\noindent
where $\mathbf{x}_{i}$, $\mathbf{v}_{i}$ and $\Phi_{i}$ (where $i = \{\mathrm{MW}, \mathrm{LMC} \}$) are the position vectors, velocity vectors and the gravitational potentials of the MW and LMC in an inertial frame, respectively. 
The extra acceleration term, $\mathbf{a}_{\mathrm{DF}}$, accounts for Chandrasekhar dynamical friction on the trajectory of the LMC \citep{Chandrasekhar1943, Binney2008, Jethwa2016},

\begin{equation}
\mathbf{a_{\rm DF}} = - \frac{4 \pi G^2 M_{\rm LMC} \rho_{\rm MW} \ln \Lambda}{v_{\rm LMC}^3} 
\left[ \operatorname{erf}(X) - \frac{2X}{\sqrt{\pi}} e^{-X^2} \right] \mathbf{v_{\rm LMC}},
\end{equation}
\noindent
where $X = v_{\rm LMC} / \sqrt{2}\sigma_{\rm MW}$ and $\sigma_{\rm MW}$ and $\rho_{\rm MW}$ are the velocity dispersion and total density field of the MW. 
Following \citet{Vasiliev2021}, we take a fixed value of $\sigma_{\rm MW} = 120 \,\rm km \, \rm s^{-1}$ for the velocity dispersion as the dynamical friction is insensitive to the precise
value. 
For the Coulomb logarithm we adopt \(\ln \Lambda = \ln \left( 100\, \rm kpc / \epsilon \right)\). The softening length, $\epsilon$, depends on the satellite’s density profile \citep{White1976}. 
We adopt $\epsilon = 1.6 \, a_{\rm LMC}$ as this has been used previously when modelling the LMC as a Plummer sphere \citep[e.g.,][]{Hashimoto2003, Besla2007, vanderMarel2012, Sohn2013, Kallivayalil2013}. The numerator in the Coulomb logarithm expression an arbitrarily chosen value that loosely describes the average separation of the MW and LMC. In principle this value could be updated through the integration of the LMC orbit, however this will have a small effect. Furthermore, our use of a dimensionless parameter, $\lambda_{\mathrm{DF}}$, that modulates the strength of the dynamical friction that the LMC experiences will take into account changes to the fixed Coulomb logarithm value per simulation. Finally, we include the acceleration due to the reflex motion of the MW's centre of mass towards the LMC.

\subsection{The Milky Way Stellar Halo}\label{sec:simulations-stellarhaloes}

To generate a mock MW stellar halo for each simulation, we draw phase-space samples from radially biased distribution functions as implemented in \texttt{agama} \citep{2019MNRAS.482.1525V}. This requires instances of a tracer density profile, a potential and a prescription for the radial velocity anisotropy. We use a Dehnen tracer density profile \citep{Dehnen1993} and an NFW profile for the potential \citep{Navarro1996, Navarro1997} with relevant parameters adopted from each unique MW--LMC simulation; see Sec.~\ref{sec:simulations-ics}.  
We implement the radial bias of stellar velocities \citep{Binney2008} using the velocity anisotropy parameter as implemented in \texttt{agama} \citep{2019MNRAS.482.1525V}:
\begin{equation}
    \beta(r) \equiv 1 - \frac{\sigma_t^2}{2\sigma_r^2} = \frac{\beta_0 + (r/r_a)^2}{1 + (r/r_a)^2}.
\end{equation}
\noindent
where $\beta_0$ is the limiting value of anisotropy in the centre, and if $r_a < \infty$, the anisotropy coefficient tends to 1 at large $r$ (Osipkov--Merritt profile), otherwise it is constant and equal to $\beta_0$ over all scales which we adopt. 
\citet{Han2024} found that $\beta$ varies as a function of distance and on-sky angle. However, other studies have found that $\beta$ takes an almost constant value throughout the Galaxy \citep[e.g.,][]{Bird2021, Cunningham2019, Chandra2025}. 
We sample the distribution function to return a set of $4500$ phase-space coordinates to represent the MW stellar halo at some early time, i.e. well before the LMC has caused any disequilibrium in the MW. We adopt this number of particles for the stellar halo as this approximately matches the average number density of stars beyond $50\,\rm kpc$ in both the \textit{Dark Energy Spectroscopic Instrument} (DESI) Blue Horizontal Branch \citep[BHB,][]{Bystrom2025} and the combined \textit{H3, + SEGUE + MagE} red giant samples \citep{Chandra2025}.
For a given MW--LMC potential with reflex motion, we integrate all particles in the stellar halo to present-day over the last $2.2\,\mathrm{Gyr}$. We choose the amount of time to evolve the MW--LMC system as it matches the live simulation time in \citet{Garavito-Camargo2019}. 

From the final distribution of stellar halo particles, we measure the reflex motion of the MW in response to the infalling LMC. To do this, we adapt the method from \citet{Petersen2021} which has subsequently been used in other works \citep[e.g.,][]{Yaaqib2024, Chandra2025}. We fit an on-sky velocity model that contains nine free parameters. We model the dipole reflex motion using Galactocentric Cartesian velocities ($v_x, v_y, v_z$). This is subtly distinct from using the magnitude of reflex motion, $v_\mathrm{travel}$, and its direction on the sphere, $(l_\mathrm{apex}, b_\mathrm{apex})$, directly as parameters \citep[][]{Petersen2021}. We do this for simplicity and to avoid inefficient convergence during the maximum likelihood estimates of these parameters when using the mock stellar halo data. The Galactocentric velocities are easily converted to their spherical equivalents by: 

\begin{gather}
    v_{\rm travel} = \sqrt{v_x^2 + v_y^2 + v_z^2}, \\
    l_{\rm apex} = \arccos{\left( v_z / \sqrt{v_x^2 + v_y^2 + v_z^2} \right)}, \\
    b_{\rm apex} = \arctan{\left(v_y / v_x\right)}
\end{gather}
\noindent
We also account for non-zero mean motion in the halo's Galactocentric velocity via the mean motion parameters $\langle v_r \rangle, \langle v_{\phi} \rangle, \langle v_{\theta} \rangle$. This allows for any departures in the bulk halo motion from the travel velocity.
Additionally, we account for the intrinsic velocity dispersion in each component using the set of hyper-parameters, $\sigma_{v_{\rm r}},\sigma_{v_l}, \sigma_{v_b}$. These dispersion parameters also absorb the measurement uncertainties.
The reflex motion model is represented by the sum of the dipole and mean motion parameters:

\begin{equation}
    \langle \vec{v}\rangle = \vec{v}_{\rm travel} + \langle v_r \rangle + \langle v_{\phi} \rangle + \langle v_{\theta} \rangle 
\end{equation}
\noindent
Where $\vec{v}$ is the mean Galactocentric halo velocity vector. We project the velocities to observable coordinates using equs.~3 \& 4 in \citet{Petersen2021}.
To find the maximum likelihood estimates for these parameters given each mock stellar halo data, we minimise a Gaussian log-likelihood for the 1-dimensional line of sight velocities and 2-dimensional proper motions using \texttt{scipy.optimize} \citep[see equs.~6 \& 8 in][]{Petersen2021}. We return the maximum likelihood estimates for all of the reflex motion model parameters. However, in the context of this work, we will only comment on $v_\mathrm{travel}, l_\mathrm{apex}, \rm{and} \, b_\mathrm{apex}$. 

\subsection{Simulation Priors}\label{sec:simulations-ics}

We summarise the prior probability distributions used for free model parameters in Table.~\ref{table:priors}.
The first three parameters are the MW mass, the infall LMC mass\footnote{Note, as the LMC mass is fixed throughout each rigid simulation, the infall and present-day LMC masses are the same.}, and dynamical friction strength, which are our target inference parameters. 
The middle set of parameters describe the LMC present-day position and velocity with their distributions inspired by the values in sec.~3.1 and table~2 of \citet{Vasiliev2023}. 
The final set of parameters are the anisotropy parameter, $\beta_0$, and the Dehnen tracer density profile scale length, $r_{\rm Dehnen}$, that initialise the mock MW stellar haloes.
In total, we run $128,000$ MW--LMC simulations with unique initial conditions for the MW and LMC as described above.

\begin{table}
\centering
\caption{Simulation model parameter prior distributions.}
\label{table:priors}
\begin{tabularx}{\linewidth}{lX}
\hline
\hline
Model Parameter & Prior probability distribution \\
\hline
& \\
\( M _{200,\rm MW} \) & \(  \mathcal{N}(15, 5) \times 10^{11} M_{\odot} \) \\
\(  M_{\rm LMC} \) & \( \mathcal{N}(15, 10) \times 10^{10} M_{\odot} \) \\
\( \log_{10}(\lambda_{\mathrm{DF}}) \) & \( \mathcal{U}(-3, 1) \) \\
 & \\
\( \alpha_{\rm LMC} \) & \(\mathcal{U} (60^\circ, 90^\circ) \) \\
\( \delta_{\rm LMC} \) & \(\mathcal{U}(-80^\circ, -50^\circ)  \) \\ 
\( d_{\rm LMC} \) & \( \mathcal{N}(49.6, 5)\,\mathrm{kpc}\) \\
\( v_{\rm los} \) & \( \mathcal{N}(262.2, 10) \,\mathrm{km}\,\mathrm{s}^{-1} \) \\
\( \mu_{\alpha_{\rm LMC}} \) & \( \mathcal{N}(1.9, 0.25)\, \mathrm{mas}\,\mathrm{yr}^{-1} \) \\ 
\( \mu_{\delta_{\rm LMC}} \) & \( \mathcal{N}(0.33, 0.25)\,\mathrm{mas}\,\mathrm{yr}^{-1} \) \\
 & \\
\(\beta_0 \) & \(  \mathcal{U}(0, 0.9)\) \\
\( r_{\rm Dehnen} \) & \(\mathcal{U}(10, 15)\, \rm kpc \) \\
\hline
\end{tabularx}
\end{table}

\section{Simulation Based Inference}\label{sec:sbi}

In a Bayesian setting, the problem is often posed as calculating the probability of the model
parameters $\theta$, given some observed data $D_{\text{obs}}$, and a theoretical model $I$. In other words, we want to find the posterior probability distribution, $\mathcal{P} = p(\theta | D_{\text{obs}}, I)$. This is possible using Bayes' Theorem:

\begin{equation}\label{equ:1}
     p(\theta | D_{\text{obs}}, I) = \frac{p(D_{\text{obs}} | \theta, I) p(\theta | I)}{p(D_{\text{obs}} | I)}
\iff \mathcal{P} = \frac{\mathcal{L} \times \Pi}{\mathcal{Z}} 
\end{equation}
\noindent
where $\mathcal{L}$ is the likelihood, $\Pi$ is the prior, and $\mathcal{Z}$ is the
Bayesian evidence. The Bayesian evidence acts as a normalisation in parameter estimation and can  be ignored for our application. Given a choice of prior distribution for parameters and a likelihood function, we can find the posterior distribution. In the case where a likelihood
function cannot be explicitly defined, or is intractable from simulation, we can use SBI to find an estimate of the posterior.

The simplest form of SBI is known as Approximate Bayesian Computation \citep[ABC, e.g.,][]{Rubin1984, Pritchard1999, Fearnhead2010}. The ABC framework selects forward simulations that are the most similar to the observed data based on some distance measure involving the summary statistics of the simulation. Another way to compute the posterior is via Density Estimation Likelihood Free Inference (DELFI). In this approach, forward simulations are used to learn a conditional density distribution of the data $D_{\text{obs}}$, given the simulation parameters $\theta$, using a density estimation algorithm, e.g., normalising flows, that utilise a series of transformations to convert a simple base distribution into the desired probability distribution \citep{JimenezRezende2015}. We use the package \texttt{sbi} \citep{tejero-cantero2020sbi}, and estimate the posterior distribution from the forward simulations using Masked Autoregressive Flows \cite[MAF,][]{Papamakarios2017, Papamakarios2019} with $5$ layers in the neural network, each with $50$ nodes. DELFI is advantageous over the simpler ABC approach as it does not rely on a choice of a distance measure and it uses all available forward simulations to build the posterior distribution, making it far more efficient \citep{Alsing2019}. Additionally, once a normalising flow has been trained on a precomputed simulation dataset, the posterior can be returned for many observations without having to retrain the flow. This is known as \textit{amortisation} \citep{Mittal2025} and is a benefit over more classical approaches to estimate the posterior such as Markov Chain Monte Carlo as the chains need to be re-evaluated every time a new observation becomes available. 

Often, some form of data compression is required \citep[e.g.,][]{Alsing2018, Alsing2019b, Heavens2020, Jeffrey2021, Widmark2025}. However, our problem has only a few parameters of interest and data points, so no data compression is required. We proceed by generating a large number of rigid MW--LMC simulations that cover broad regions of parameter space. Generally, for SBI, the more simulations that are available to use, the better. Recently, \citet{Bairagi2025} estimated the number of simulations that are required, $\sim\mathcal{O}(10^{3}-10^{4})$, when using SBI within a cosmological context. The simulations used in this work are described in Sec.~\ref{sec:simulations}. 
We will then use a MAF density estimator from the \texttt{sbi} package \citep{tejero-cantero2020sbi} to directly obtain the posterior distribution that can be evaluated at any observed data point for any data realisation, i.e, \( p(\theta | D_{\text{obs}}, I) \). In Sec.~\ref{sec:posterior-diagnostic-checks}, we provide a complete discussion of posterior probability diagnostic checks.

\section{Results}\label{sec:validation}

We perform a series of tests to assess the validity of our SBI framework.
First, in Sec.~\ref{sec:validation-test-rigid} , we select a random simulation from our large set of rigid MW--LMC simulations. In this closed-loop test, we aim to recover some of its true simulation parameters by adopting some of the other simulation parameters as ‘data points'. 
Second, in Sec.~\ref{sec:validation-test-gc21}, we extend this concept to recover simulation parameters from well-known idealised deforming MW--LMC models \citep{Garavito-Camargo2019, Garavito-Camargo2021}. 
Finally, in Sec.~\ref{sec:validation-test-cosmosims}, we push this formalism further to a MW--LMC analogue pair identified in the FIRE latte suite of \citep{Wetzel2016} cosmological simulations.
In all figures we use $10, 000$ samples drawn from their respective posterior density
distributions.

\subsection{On rigid MW--LMC models}\label{sec:validation-test-rigid}

\begin{figure*}
    \centering
    \begin{overpic}[width=1\linewidth]{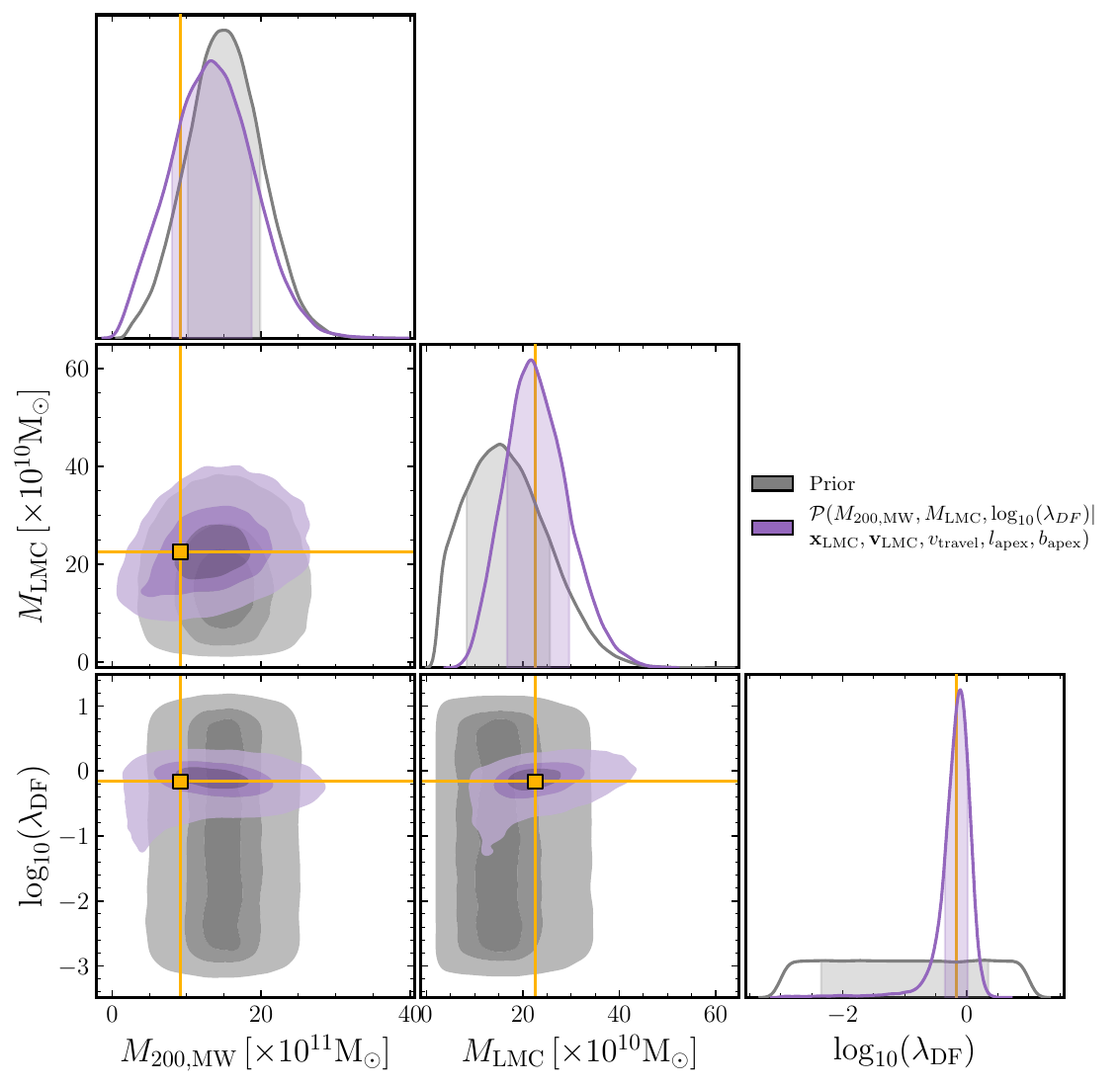}
         \put(52,65){\includegraphics[width=0.5\linewidth]{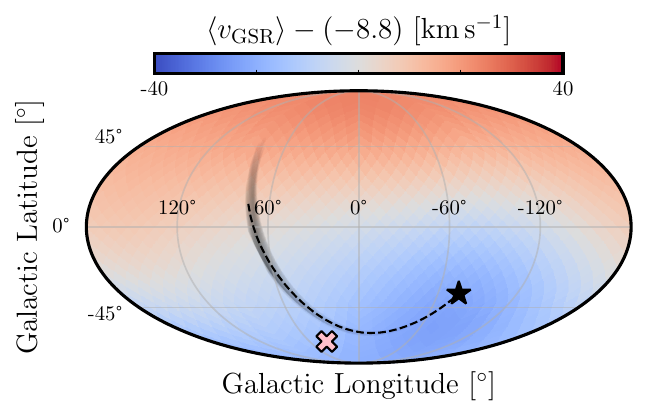}}  
      \end{overpic}
    \caption{\textbf{Validation on a rigid MW--LMC model:} The joint, and individual, posterior distributions for MW mass, LMC mass and dynamical friction strength when there is knowledge of the LMC present-day position, velocity and the reflex motion parameters (purple). We also show the parameter prior distributions in grey. For the 1D posterior panels we show the $16^{\rm th}$ and $84^{\rm th}$ percentiles as shaded regions. The yellow square/lines represent the true values from the randomly selected rigid MW--LMC simulation from our larger simulation sample. The reflex motion parameters $v_\mathrm{travel}, l_\mathrm{apex}, b_\mathrm{apex}$ break model parameter degeneracies to add constraining power, particularly in the LMC mass and dynamical friction strength. The all-sky projection displays the average solar corrected radial velocity, $v_{\rm GSR}$, for all stellar halo particles beyond $50\,\rm kpc$. The net signal across the entire sky is subtracted. The pink cross denotes the reflex motion apex for this simulation. The past orbit of the LMC in this potential is shown as the black dashed line. Orbits computed using posterior samples for the masses and dynamical friction strength, for fixed LMC present-day conditions, are shown as the grey shaded region.}
    \label{fig:1}
\end{figure*}

To begin with a simple example, we aim to recover true parameters of a rigid MW--LMC simulation that uses our prescription for the potential as outlined in Sec.~\ref{sec:simulations}. By selecting a simulation from our sample, which is not used to train the SBI framework, we expect to be able to find posterior distributions which enclose the true parameters of the simulation.
The randomly chosen simulation has parameter values as follows:
The LMC has a present-day Galactocentric position, $\mathbf{x}_{\mathrm{LMC}} = \{-1.1,  -28.0,  -21.5\} \, \rm kpc $, and velocity, $\mathbf{v}_{\mathrm{LMC}} = \{-87.1, -198.9 ,  135.4\} \, \rm km \, \rm s^{-1}$. 
The LMC mass is $\rm M_{LMC} = 23 \times 10^{10}\,\rm M_{\odot}$, with scale radius, $a_{\rm LMC} = 18.9\, \rm kpc$.
The MW mass is $\rm M_{200, MW} = 9.1 \times 10^{11}\,\rm M_{\odot}$, with concentration $c_{200} = 14.5$. 
For the stellar halo, the anisotropy parameter is $\beta_0 = 0.76$, and the Dehnen scale radius is $11.5\, \rm kpc$ for the density profile. 
In this MW--LMC potential, the MW reflex parameters as measured from the stellar halo are $v_\mathrm{travel} = 64.5 \,\rm km \, \rm s^{-1}, l_\mathrm{apex} =  39.1^{\circ}, b_\mathrm{apex} = -67.8^{\circ}$; see Sec.~\ref{sec:simulations-stellarhaloes}. 
The dynamical friction scalar strength used is $\lambda_{\rm DF} = 0.7$. 

We aim to recover the MW mass, LMC mass, and dynamical friction strength assuming we have knowledge of the LMC present-day position, velocity, and the reflex motion parameters. In this case, the Bayesian notation in Equ.~\ref{equ:1} follows as: 
\begin{gather*}
    \theta = \{\rm M_{LMC}, \rm M_{200, MW}, \log_{10}(\lambda_{\rm{DF}}) \}, \\
        D_{\text{obs}} = \{ \mathbf{x}_{\mathrm{LMC}}, \mathbf{v}_{\mathrm{LMC}}, v_\mathrm{travel}, l_\mathrm{apex}, b_\mathrm{apex} \}    
\end{gather*}
\noindent
In Fig.~\ref{fig:1} we show the joint, and individual, posterior distributions for the LMC and MW total masses using a set of $128,000$ MW--LMC simulations.
The true values for the MW mass, LMC mass, and dynamical friction strength (yellow square and lines) all lie within the $16^{\rm th}$ and $84^{\rm th}$ percentiles of the posterior (purple shaded region in 1D insets). 
We additionally show the non-normalised prior distributions in grey to give intuition about the constraining power in each parameter. 
The LMC mass and dynamical friction strength are both accurately and precisely constrained by the supplied data values for the LMC present-day position, velocity, and the reflex motion parameters. On the other hand, the MW mass is accurately constrained, although its spread, i.e., its precision, remains similar to the original prior. This suggests that the learned MW mass posterior is somewhat insensitive to the given data points. 
Much of the constraining power for the LMC comes from providing the data points for $v_{\rm travel}, l_{\rm apex}, b_{\rm apex}$; see the all-sky projection for the direction of the apex on-sky (pink cross). 
This is not unexpected as the LMC mass and the magnitude of the travel velocity, $v_{\rm travel}$, are positively correlated \citep[e.g.,][]{Petersen2021}. 
This implies that providing $v_{\rm travel}$ as a data point is crucial in breaking degeneracies for the returned posteriors.
Indeed, we have performed inference on the same model parameters but only conditioning on the LMC present day position and velocity. 
In this case, we find that the posterior closely resembles the prior, implying that most of the constraining power comes from the reflex motion measurements.

\subsection{On deforming MW--LMC models}\label{sec:validation-test-gc21}

\begin{figure*}
    \centering
     \begin{overpic}[width=1\linewidth]{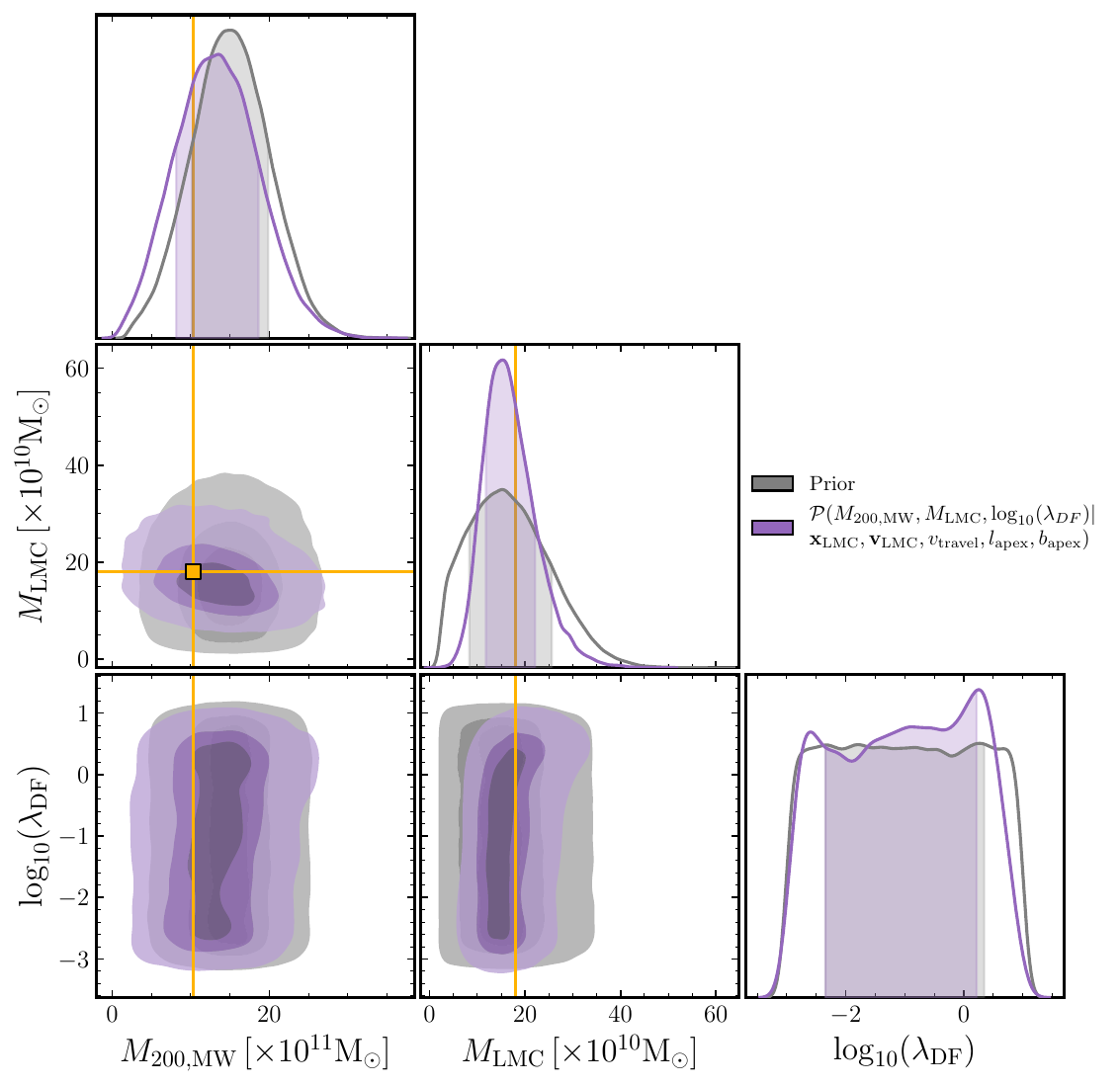}
         \put(52,65){\includegraphics[width=0.5\linewidth]{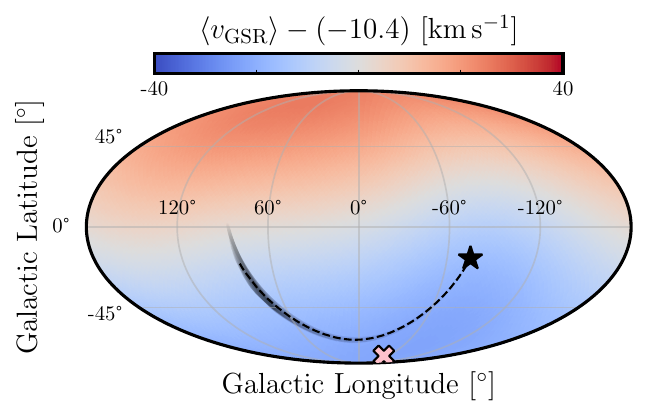}}  
      \end{overpic}
    \caption{\textbf{Validation on a deforming MW--LMC models:} The joint, and individual, posterior distributions for MW mass, LMC mass and dynamical friction strength when there is knowledge of the LMC present day position, velocity and the reflex motion parameters (purple). We also show the parameter prior distributions in grey. For the 1D posterior panels we show the $16^{\rm th}$ and $84^{\rm th}$ percentiles as shaded regions. The yellow squares/lines represent the true values from the fiducial deforming simulation of \citet{Garavito-Camargo2019}. The reflex motion parameters  $ v_\mathrm{travel}, l_\mathrm{apex}, b_\mathrm{apex}$ break model parameter degeneracies to add constraining power, particularly in the LMC mass. The all-sky projection displays the average solar corrected radial velocity, $v_{\rm GSR}$, for all stellar halo particles beyond $50\,\rm kpc$. The net signal across the entire sky is subtracted. The pink cross denotes the reflex motion apex for this simulation. The past orbit of the LMC in this potential is shown as the black dashed line. Orbits computed using posterior samples for the masses and dynamical friction strength, for fixed LMC present-day conditions, are shown as the grey shaded region.}
    \label{fig2}
\end{figure*}

We assess the validity of our sample of rigid MW--LMC simulations to be able to confidently capture the dynamics of the more complicated deforming MW--LMC simulations. In other words, we want to assess whether the rigid set-up can be upscaled in their fidelity to resemble the more complex, deforming MW--LMC simulations.
In particular, we aim to recover the true parameters from the \citet{Garavito-Camargo2019} deforming MW--LMC simulations. 
We choose to recover parameters from the fiducial simulation run. 
In this simulation, the LMC has a present-day Galactocentric position, $\mathbf{x}_{\mathrm{LMC}} = \{-2.32, -45.57, -21.45\} \, \rm kpc $ and velocity $\mathbf{v}_{\mathrm{LMC}} = \{-59.9, -249.3, 227.9\} \, \rm km \, \rm s^{-1}$.
The fiducial infall LMC mass is $M_{\rm LMC} = 18 \times 10^{10}\,\rm M_{\odot}$ with scale radius $a_{\rm LMC} = 20\, \rm kpc$. 
The fiducial MW mass is $M_{\rm 200, MW} = 10.3 \times 10^{11}\,\rm M_{\odot}$ with concentration $c_{200} = 11.0$. 
To measure the MW reflex parameters, we take all of the MW halo dark matter particles from the simulation and apply the same particle selection criteria as in Sec.~\ref{sec:simulations-stellarhaloes} for our mock stellar haloes. This ensures that the number of particles approximately matches the average number density of stars beyond $50\,\rm kpc$ in the DESI BHB sample within the DESI sky-coverage footprint \citep[][adopting equ.~11 for the DESI footprint]{Bystrom2025}. The resulting reflex motion parameters as measured from this simulation are $v_\mathrm{travel} = 27.9\,\mathrm{km} \, \mathrm{s}^{-1}, l_\mathrm{apex} = -49.1^{\circ}, b_\mathrm{apex} = -80.4^{\circ}$; see the all-sky projection in Fig.~\ref{fig2} for the direction of the apex on-sky (pink cross). 

Once again, we aim to recover the MW mass, LMC mass, and dynamical friction strength evaluated using the same simulation data points i.e., supplying knowledge of the LMC present day position, velocity and the MW reflex motion parameters. As our posterior is amortised, this is a trivial exercise using the same learned posteriors as in Sec.~\ref{sec:validation-test-rigid}.
In Fig.~\ref{fig2} we show the joint, and individual, posterior distributions. Encouragingly, our SBI framework is able to recover the true LMC mass from the deforming MW--LMC simulations whilst only using rigid models. 
However, very little constraining power is gained for the MW mass and dynamical friction strength. 
No specific value for the ‘true' dynamical friction strength is quoted for this deforming simulation, hence its absence from Fig.~\ref{fig2}. Although, one can use the LMC orbit in this simulation to find an approximate value given our treatment of dynamical friction, see Sec.~\ref{sec:simulations-mwlmcinteraction}, which returns a value consistent with fiducial Chandrasekhar dynamical value i.e., $\log_{10}(\lambda_{\rm DF}) = 0$. 
In an attempt to break model parameter degeneracies to recover such a value would require even more data to be provided. 

\begin{figure}
    \centering
    \includegraphics[width=\linewidth]{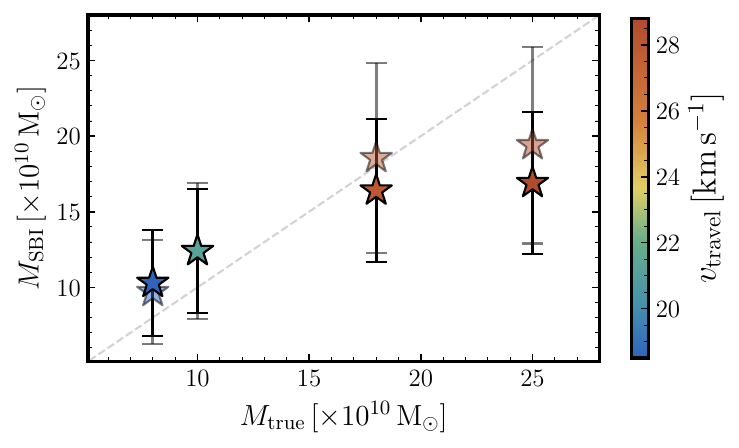}
    \caption{\textbf{Comparison of the true and inferred infall LMC masses:} We compare the simulation truth and returned SBI posterior LMC mass for all of the MW--LMC deforming simulations in \citet{Garavito-Camargo2019}. 
    We show the $16^{\rm th} - 84^{\rm th}$ percentiles of the LMC mass posteriors as error bars. For all deforming MW--LMC simulations except the most massive LMC, the true and SBI returned LMC masses are consistent, i.e., they overlap on the 1:1 dashed grey line, within the $1\,\sigma$ confidence interval. Additionally, we show the re-weighted SBI masses, which account for any biases in the choice of the LMC mass prior, as the fainter points and errors. We colour each point by the measured simulation travel velocity to demonstrate its expected increase with increasing LMC mass.}
    \label{fig3}
\end{figure}

Motivated by the LMC mass posterior in Fig.~\ref{fig2} for the fiducial deforming MW--LMC simulation of \citet{Garavito-Camargo2019}, we also take the other deforming simulations data from this suite that adopt the same MW model but have varying LMC masses i.e., $\mathrm{M_{LMC}} = \{8, 10,\, 25\} \times 10^{10}\,\mathrm{M}_{\odot}$. The LMC masses are defined to be their mass when each LMC first crossed the virial radius of the MW. Repeating the above inference methodology with these deforming MW--LMC simulations, we obtain the LMC mass posterior distribution when assuming knowledge of the LMC present day position, velocity, and reflex motion parameters. We compare the true simulation LMC mass to the median, plus $1\,\sigma$ confidence interval, of the SBI LMC mass posteriors in Fig.~\ref{fig3}. For all except the most massive LMC model in \citet{Garavito-Camargo2019}, we find that the true and SBI LMC masses are consistent, i.e., they overlap on the 1:1 grey line, within the $1\,\sigma$ confidence interval. 
To explain this, we re-weight the returned SBI median and percentiles to account for sampling the LMC mass from a Gaussian prior probability distribution and instead adopt a flat prior distribution (fainter points and error bars). 
This accounts for any biases that are introduced as a result of the choice of the Gaussian prior distribution.
In Sec.~\ref{sec:discussion-priors} we discuss further the effect of the choice of prior. 
Now, all of the true and SBI LMC masses are consistent within their $1\,\sigma$ confidence intervals. Intuitively this agrees with expectation, as the most massive LMC, $M_{\rm LMC} = 25\times 10^{10}\,\mathrm{M}_{\odot}$ is towards the lower probability tails of the adopted Gaussian prior distribution e.g., see grey contours in Fig.~\ref{fig2}. Hence, any SBI estimate of the LMC mass would be expected to be biased to lower values based on the choice of our prior. Conversely, the opposite is also true for the least massive LMC; see Fig.~\ref{fig3}.  
Furthermore, we demonstrate the expected increase in measured travel velocity (marker fill colour) as the LMC mass is increased \citep{Petersen2021}. Overall, our SBI framework is able to correctly recover the LMC mass well from more complex, deforming MW--LMC simulations. This is encouraging for future applications of this technique to constrain the real LMC mass using observational data \citep[this has been performed in a simultaneous and distinct study, see][]{Brooks2025}.

\subsection{On a cosmological MW--LMC analogue}\label{sec:validation-test-cosmosims}

\begin{figure*}
      \centering   
      \begin{overpic}[width=1\linewidth]{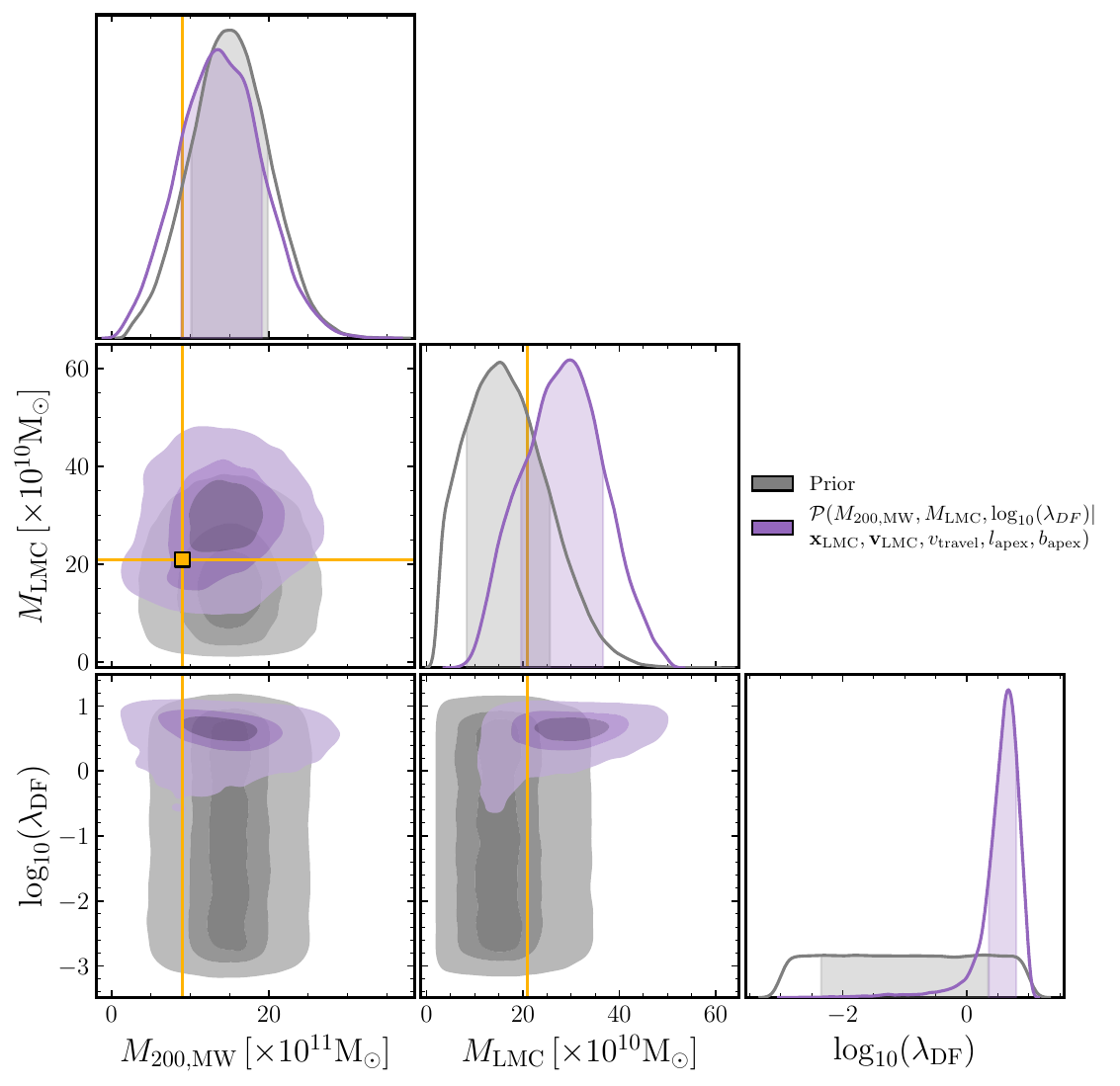}
         \put(52,65){\includegraphics[width=0.5\linewidth]{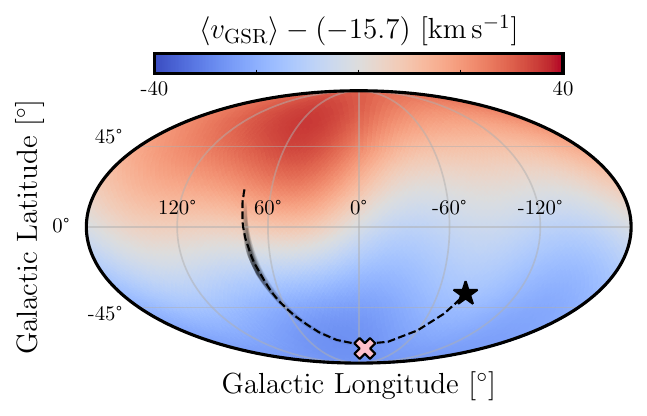}}  
      \end{overpic}
    \caption{\textbf{Validation on a cosmological MW--LMC analogue:} The joint, and individual, posterior distributions for MW mass, LMC mass and dynamical friction strength when there is knowledge of the LMC present day position, velocity and the reflex motion parameters (purple). We also show the parameter prior distributions in grey. For the 1D posterior panels we show the $16^{\rm th}$ and $84^{\rm th}$ percentiles as shaded regions. The yellow squares/lines represent the true values of \textit{m12b}'s MW and LMC analogue from the FIRE latte project \citep{Wetzel2023}. The all-sky projection displays the average solar corrected radial velocity, $v_{\rm GSR}$, for all stellar halo particles beyond $50\,\rm kpc$. The net signal across the entire sky is subtracted. The pink cross denotes the reflex motion apex for this simulation. The past orbit of the LMC in this potential is shown as the black dashed line. Orbits computed using posterior samples for the masses and dynamical friction strength, for fixed LMC present-day conditions, are shown as the grey shaded region.}
    \label{fig4}
\end{figure*}

In this section, we aim to push our SBI formalism to capture the complexity of a cosmological simulation analogue of the MW--LMC system. Our simple, rigid MW--LMC simulations cannot capture the complex physical processes of a hydrodynamical cosmological simulation. Nevertheless, we look to use only the dynamics of the MW--LMC interaction to constrain their properties in this setting.

To demonstrate this, we use a MW analogue from the \textit{Latte} suite within the FIRE \citep{Wetzel2023} project\footnote{\href{https://fire.northwestern.edu/latte/}{Feedback In Realistic Environments webpage}}. These simulations are zoom-in cosmological-hydrodynamical simulations run using the FIRE-2 model \citep[see,][for a detailed description of the full physics model]{Hopkins2018}. We select the \texttt{m12b} simulation from the \textit{Latte} suite as this is the closest analogue to the real MW-–LMC system \citep[see,][table~1 for a summary of the \textit{Latte} MW--LMC analogue properties]{Garavito-Camargo2024}. This simulation produces a present-day MW with a stellar mass, gas mass and the dark matter mass / density profile that are similar to the real MW \citep{Hopkins2018, Garrison-Kimmel2018, Sanderson2020}. The present-day mass of the MW analogue is $M_{200, \rm MW} = 9.0\times 10^{11}\,\rm M_{\odot}$. Meanwhile, the LMC analogue in \texttt{m12b} has an infall mass of $ M_{\rm LMC} = 21 \times 10^{10}\,\rm M_{\odot}$ such that its mass ratio to the MW host is $\sim1:4$. 

The \textit{Latte} suite of simulations are carried out in an arbitrary simulation box frame. However, in order to take advantage of our rigid MW--LMC simulation set described in Sec.~\ref{sec:simulations}, we must rotate this cosmological analogue to have the same orientation as the real MW--LMC system. To achieve this, we follow the same procedure as in appendix A of \citet{Arora2024} whereby we first establish a principal axis that aligns the galactic disc to the $x-y$ plane at the epoch of the analogue’s first pericentric passage. Thereafter, an additional rotated axis is defined to align the position of the LMC analogue at the epoch of first pericentric passage with that of the real LMC. To do this, the real pericentre of the LMC is based on the orbits in \citet{Garavito-Camargo2019}. Applying these rotational transformations to \texttt{m12b}'s LMC analogue gives a present-day position of $\mathbf{x}_{\mathrm{LMC}} = \{1.8,  -16.7,  -34.0\} \, \rm kpc $ and a velocity of $\mathbf{v}_{\mathrm{LMC}} = \{ -87.7, -313.4,  149.4\} \, \rm km \, \rm s^{-1}$.
The same rotational transformations are applied to the stars of the \texttt{m12b} MW analogue. In this frame, we can measure the MW reflex parameters using the method outlined in Sec.~\ref{sec:simulations-stellarhaloes}.
The stars of the \texttt{m12b} MW analogue have been masked for any substructures that could lead to biasing of the returned parameters. In this context, substructure is defined as any stars that are associated with satellite galaxies and the LMC analogue.
Again, we ensure that the number of particles approximately matches the average number density of stars beyond $50\,\rm kpc$ in the DESI BHB sample within the DESI sky-coverage footprint \citep[][adopting equ.~11 for the DESI footprint]{Bystrom2025}. 
The resulting reflex motion parameters as measured from this simulation are $v_\mathrm{travel} = 50.0 \,\rm km \, \rm s^{-1}, l_\mathrm{apex} = -8.97^{\circ}, b_\mathrm{apex} = -73.4^{\circ}$; see the all-sky projection in Fig.~\ref{fig4} for the direction of the apex on-sky (pink cross).

We look to estimate the MW analogue mass, LMC analogue mass and the dynamical friction strength with knowledge of the LMC present-day position, velocity and the reflex motion parameters. Using our amortised posterior this can be efficiently evaluated for the given data. 
In Fig.~\ref{fig4} we show the joint, and individual, posterior distributions. Interestingly, our SBI framework is able to recover the true LMC mass from the cosmological MW--LMC analogue within the $16^{\rm th}$ and $84^{\rm th}$ percentiles. However, the mean of the LMC mass posterior distributions is biased towards higher values. 
Moreover, the dynamical friction strength is well constrained to values that are $\sim 2-3$ times greater than, and intriguingly not consistent with, those expected by the classical Chandrasekhar dynamical friction \citep[i.e., $\log_{10}(\lambda_{\rm DF}) = 0$,][]{Chandrasekhar1943}. Similar to the deforming simulations, no specific value for the ‘true' dynamical friction strength is quoted for this cosmological simulation, hence its absence from Fig.~\ref{fig4}. Again we can use the orbit of this LMC analogue to find an approximate value given our treatment of dynamical friction which returns a value consistent with the posterior in Fig.~\ref{fig4}. Exploring the ability to use our SBI framework to constrain dynamical friction remains of interest for future studies. 
Once more, the MW mass posterior is not well constrained away from the original prior distribution, implying that the supplied data points do not break model degeneracies. 

\section{Posterior Diagnostic Tests}\label{sec:posterior-diagnostic-checks}

\subsection{Coverage probability test}

Any posterior resulting from a generative model should be assessed for its accuracy through a variety of diagnostic tests before being used for inference given the actual observed data. The accuracy of an estimated posterior is often performed using some form of a coverage probability test. 
A coverage test in Bayesian analysis checks whether credible intervals have the expected probabilities \citep[see,][sec.~2.4 for a concise explanation]{Jeffrey2025}. 
In a 1-dimensional posterior setting, one can define a particular credible interval to be
the narrowest interval containing, for example, $90\%$ of the probability weight.
The Bayesian inference procedure takes in some observed data, $D_{\rm obs}$, and determines a posterior
distribution, $p(\theta | D_{\rm obs})$, and hence a credible interval for $\theta$. 
For a coverage test, one uses a test parameter, $\theta_{\rm test}$, selected from the prior, $p(\theta)$, as the input to a simulation that produces the corresponding output data point, $D_{\rm test}$. From this, one can derive a posterior, $p(\theta | D_{\rm test})$, and therefore a credible interval. If the inference process has been correctly implemented then the true test parameter value, $\theta_{\rm test}$, will fall in this credible interval, in this example, $90\%$ of the time. 
Repeating this test for many sampled $\theta_{\rm test}$, and varying confidence intervals, one can gain confidence that the estimated posterior distributions are indeed correct.

\begin{figure}
    \centering
    \includegraphics[width=\linewidth]{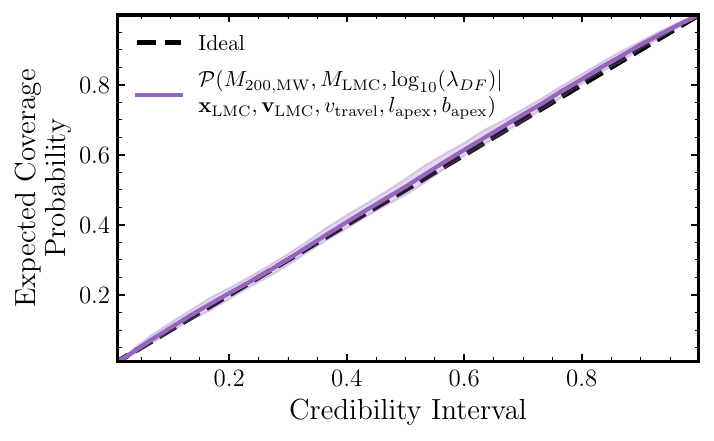}
    \caption{\textbf{Tests of Accuracy with Random Points:} This diagnostic check acts as a coverage probability test. The probability of true values in the appropriate credible intervals matches the expected coverage probability. A bootstrapped $2\sigma$ uncertainty is shown as the shaded region. This validates our SBI posterior estimation such that the posterior is representative of the probability that each of our model parameters has some true value. }
    \label{fig5}
\end{figure}

To perform a coverage test on our SBI posteriors, we will follow the ‘Tests of Accuracy with Random Points' \citep[TARP,][see their figs.~1\&2 for futher intuition]{Lemos2023}. For our application of SBI, this test is relatively straightforward as we have many pre-existing simulations with an amortised inference scheme, i.e., each data evaluation is computationally cheap.
TARP coverage probabilities test the accuracy of estimated posteriors by only using samples made from the posterior. This technique is similar to simulation based calibration \citep{Talts2018} but extends the idea to the full-dimensional posterior space instead of being solely 1-dimensional. We use the implementation of TARP in the \texttt{sbi} python package \citep{tejero-cantero2020sbi} and show our results in Fig.~\ref{fig5}.

A valid SBI posterior will have a coverage probability that is similar to any given credibility interval one chooses. Fig.~\ref{fig5} demonstrates that the expected coverage does indeed match the credibility level. This validates our neural posterior estimation as being truly representative of the probability that each of our model parameters has some true value.
This can be further quantified in two ways. 
Firstly, we can compute the area between the ideal TARP curve and our posterior TARP curves for credibility intervals greater than $0.5$; namely the Area To Curve (ATC) value. This number should be close to $0$, a value $\gg0$ indicates an estimated posterior that is too wide, conversely, a value $\ll0$ indicates that the estimated posterior is too narrow. 
Secondly, we can calculate the p-value of a Kolmogorov-Smirnov test. The null hypothesis is that the ideal TARP curve and our posterior TARP curve are identical. If this p-value is less than $0.05$, then this null hypothesis is rejected. 
For our posterior distributions of MW mass, LMC mass, and dynamical friction strength when there is knowledge of the LMC present-day position, velocity, and reflex motion parameters (purple) we report an ATC value of $0.27$ and a Kolmogorov-Smirnov p-value of $1.0$. These values suggest that we are not drastically over-/under-fitting and are not required to reject the posterior.

\subsection{Posterior Predictive Check}

Although not explicitly a diagnostic test, we carry out a \textit{‘Posterior Predictive Check'} (PPC) to act as a complementary diagnostic test. This test makes use of the fact that if the inference is correct, then any generated data, $D_{\rm pp}$, using simulation parameter values sampled from the posterior, $\theta_{\rm pp}$, should be similar to the observed data, $D_{\rm obs}$ \citep[][]{Lueckmann21a}. A PPC can provide an intuition about any bias introduced in inference e.g., determining whether or not the generated data systematically differ from the observed data used during the inference. 

\begin{figure}
    \centering
    \includegraphics[width=\linewidth]{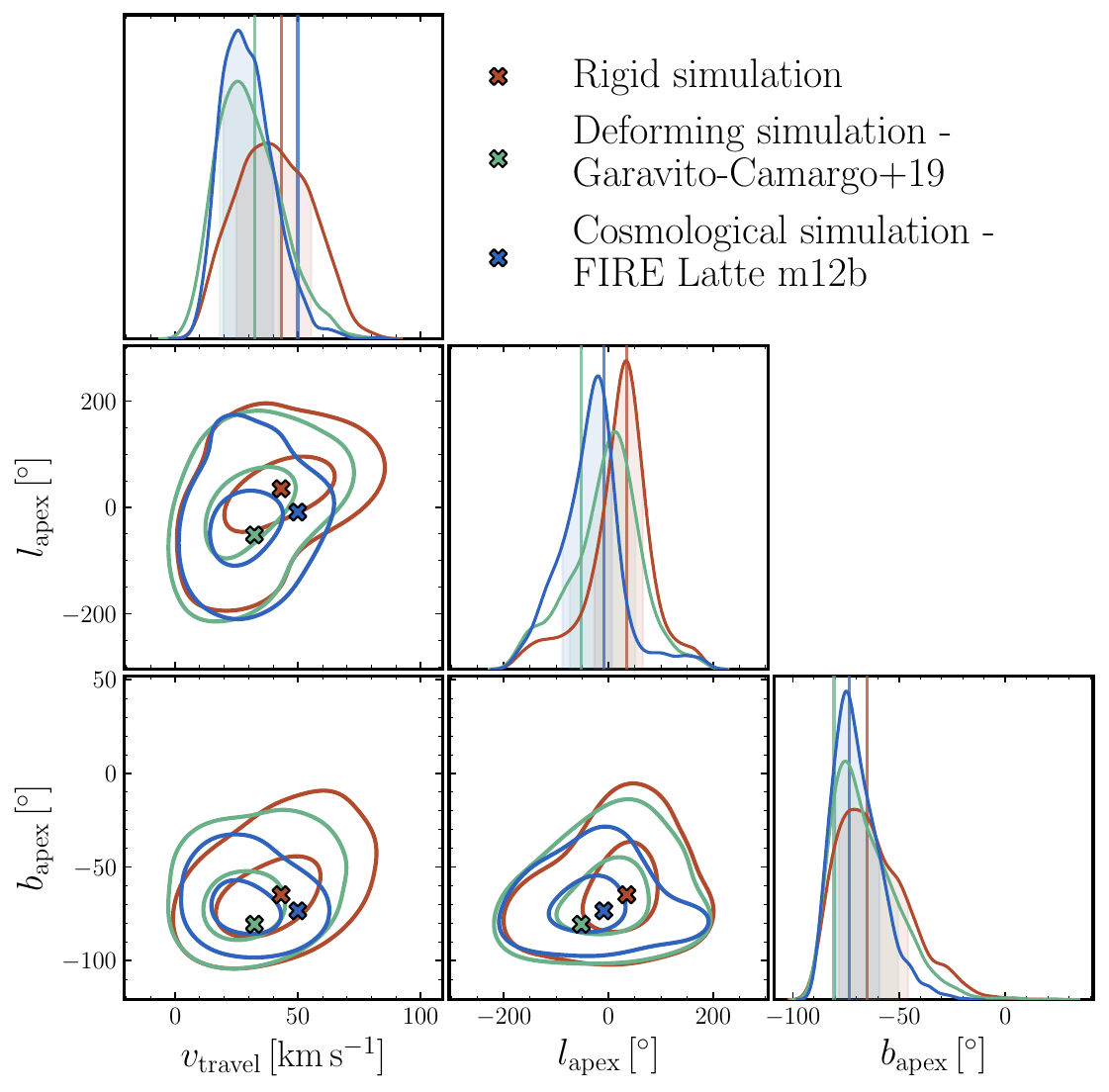}
    \caption{\textbf{Predictive Posterior Check:} This diagnostic test assesses whether a posterior learned using a SBI approach is representative of the observed data. In particular, we assess the quality of the returned posterior distributions for MW mass, LMC mass and dynamical friction strength when there is knowledge of the LMC present day position, velocity and knowledge of the reflex motion parameters. We compare the generated data contours using posterior samples from a rigid simulation (red), the deforming simulation (green) and the cosmological simulation (blue) with the original data used to evaluate the posteriors as the same coloured crosses/vertical lines. The generated and original data look sufficiently similar such that we are confident that no biases are introduced during the SBI analysis.}
    \label{fig6}
\end{figure}

We carry out a PPC for $1,024$ samples from our posterior distributions of the MW mass, LMC mass, and dynamical friction strength when there is knowledge of the LMC present-day position, velocity, and knowledge of the reflex motion parameters. We repeat this for the three cases we consider in this work:~the validation on a rigid simulation in Sec.~\ref{sec:validation-test-rigid}, the validation on a deforming simulation in Sec.~\ref{sec:validation-test-gc21} and the validation on a cosmological simulation in Sec.~\ref{sec:validation-test-cosmosims}. To do this, we sample values for the MW mass, LMC mass and dynamical friction strength from the posterior evaluated using data in these three scenarios. We then re-run rigid MW--LMC simulations adopting these parameter values and output the reflex motion parameters to compare to the original input reflex motion data. 
In Fig.~\ref{fig6}, we visualise the resulting generated data as coloured $1\,\sigma$ and $2\,\sigma$ contours and the original data (coloured crosses). The 1D panels show the filled $1\,\sigma$ region. Clearly, the input reflex motion data points used to evaluate the posteriors fall well within the $2\,\sigma$ confidence intervals across all fidelity MW--LMC simulations we consider. Furthermore, for the rigid and deforming simulations, the input data is contained within their $1\,\sigma$ confidence interval. This suggests that our return SBI posteriors are not subject to any biases and are truly representative of the input data. 

\subsection{Train--Test split}

\begin{figure}
    \centering
    \includegraphics[width=\linewidth]{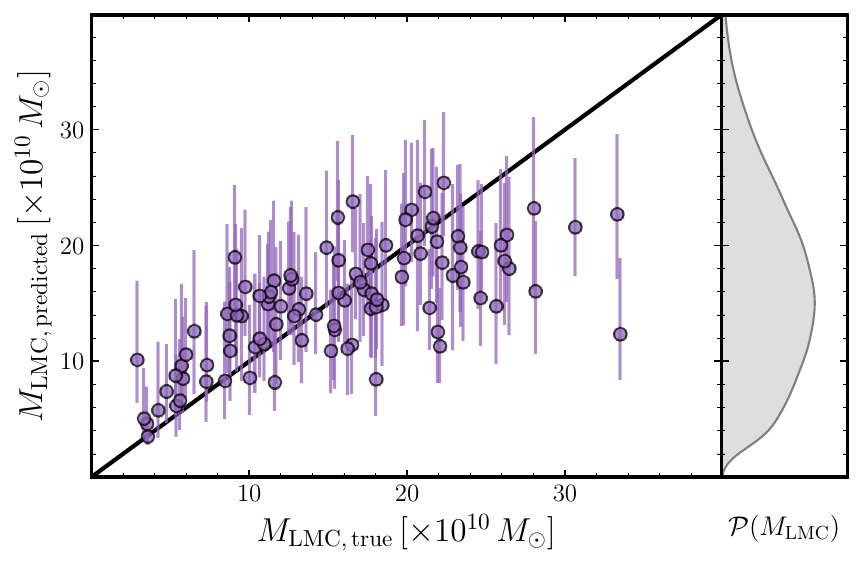}
    \caption{
    \textbf{True vs.~predicted LMC mass for a test set of simulations:} we show the true LMC mass compared to its predicted median value, with $16^{\rm th}-84^{\rm th}$ percentile error bars, for a test set of $100$ of our rigid simulations which have not been used to train the neural posterior estimator. 
    The training set used to learn the amortised posterior distribution consists of all other simulations.
    Towards the largest LMC masses, the neural posterior estimator biases the estimated LMC mass towards lower values, driven by the prior probability coverage; see right panel.
    }
    \label{fig7}
\end{figure}

We carry out a test--train split check to determine whether the posterior is performing accurately across on the full parameter ranges.
To achieve this, we reserve $5,000$ of our rigid simulations to act as the test set which are not used to train the neural posterior estimator.
The posterior training set consists of all other simulations.
In Fig.~\ref{fig7}, we compare the true LMC mass to its predicted median value, with $16^{\rm th}-84^{\rm th}$ percentile error bars, for a random sample of $100$ rigid simulations, taken from the larger test set of $5,000$ simulations. 
The true and predicted LMC masses agree within uncertainty except for the very largest LMC masses, where the prior begins to bias the predicted mass to lower values. 
Nonetheless, this high mass region is unlikely to contain the true present-day LMC mass, making this deviation inconsequential.

\section{Discussion}\label{sec:discussion}

\subsection{Caveats}\label{sec:discussion-caveats}

\subsubsection{Improvements to the Milky Way -- LMC models}

Firstly, we have assumed that the MW is at equilibrium prior to the LMC's infall. The extent to which the MW has any ‘smooth’ stellar halo is still of interest.
The inner halo is made up of many substructures from accreted dwarf galaxies, globular clusters and stellar streams, some of which are well-mixed in phase-space, whereas others retain a degree of coherent structure at present day \citep[e.g.,][]{Naidu2020, Malhan2022, Dillamore2025}. 
Similarly, the outer halo contains a plethora or accreted substructures \citep[e.g., the merger with the Sagittarius dSph galaxy,][and references therein]{Helmi2020, Vasiliev2021, Hunt2025}. 
Accounting for this induced state of disequilibrium in the MW prior to the LMC's infall is an oversight in our current modelling. 
One way to incorporate this is to include an overall net compression velocity to the MW stellar halo due to its contraction over time \citep{Chandra2025, Bystrom2025}. 
Additionally, we are not required to account for masking substructures that are seen in the observed MW stellar halo \citep[e.g., the Sagittarius stellar stream,][]{Amarante2024, Bystrom2025}.

Secondly, we assumed that the LMC has recently made its first pericentric passage. However, there is the alternative scenario where the LMC is on its second pericentric passage \citep{Vasiliev2024}. In this case, the first pericentric passage would cause a degree of disequilibrium in the MW, which we do not account for. However, the first pericentric passage would have occurred around $\sim 5 - 10 \,\rm Gyr$ ago and at a distance $\geq 100\,\rm kpc$. Hence, almost all dynamical disequilibrium induced by any previous passage will have been superseded by its most recent one \citep{Vasiliev2024, Sheng2024}.

Other modelling improvements include accounting for any mass changes of either the MW or LMC, triaxiality in the MW stellar haloes and/or distortions. 
The mass changes are especially important when considering cosmological analogues as both the host and satellite masses can change by a significant fraction throughout their interaction. In turn, this affects the degree of dynamical friction that the infalling satellite is subject to. In our current analytic dynamical friction prescription, this makes it difficult to separate out the effect of a time varying satellite mass and the inferred value of the dynamical friction strength at present-day, $\lambda_{\rm DF}$ \citep[see Sec.~\ref{sec:simulations-mwlmcinteraction} and][]{Koposov2023}.
A further future model improvement would be to implement a more careful treatment of time dependent dynamical friction to account for this. Moreover, Chandrasekhar's formula becomes increasingly ill-posed when considering two body systems of comparable masses \citep{White1983, Binney2008, Jethwa2016}. 
Finally, $v_{\rm travel}$ is sensitive to the mass enclosed by the LMC within a given distance rather than the total LMC mass which involves an extrapolation. We aim to reflect this in future work to return posteriors on the LMC mass enclosed. 

\subsection{Choice of prior distributions}\label{sec:discussion-priors}

To assess whether our inference of the MW and LMC masses are being biased by the choice of informative prior, we run a distinct set of $4,096$ simulations with the MW and LMC masses drawn from uninformative, flat, prior distributions. These are a distinct sample of simulations from those used throughout the main body of this work and are only  used in this section. Specifically we sample the masses from \( M_{200,\rm MW} \sim \,\mathcal{U}(1, 100) \times 10^{11} M_{\odot} \) and \( M_{\rm LMC} \sim \, \mathcal{U}(2.5, 100) \times 10^{10} M_{\odot} \).
We then randomly select one of these rigid simulations from our sample and we aim to recover its true parameters. 
This test exactly follows the methodology of Sec.~\ref{sec:validation-test-rigid}. 
The only difference being that the MW and LMC masses were drawn from uninformative rather than informative prior distributions. 
As before, we estimate the MW and LMC mass posterior distributions with knowledge of the LMC present day position and velocity and MW reflex motion parameters. 
We find that the true simulation mass values fall well within the $16^{\rm th} - 84^{\rm th}$ percentiles of the posterior. 
This indicates that, in the case where we supply reflex motion information as data points, our SBI framework is not subject to any confirmation biasing of the MW and LMC mass by using informative prior distributions. 

\subsection{Cosmological MW--LMC analogues}\label{sec:discussion-cosmosims}

Our validation test to use the presented SBI formalism to capture the complexity of a cosmological simulation analogue of the MW--LMC system in Sec.~\ref{sec:validation-test-cosmosims} focused only on a single analogue; FIRE Latte \texttt{m12b}. Indeed, many other MW--LMC analogues have been identified within cosmological simulations e.g., there are more within the FIRE Latte suite itself \citep{Wetzel2023}, plus many more in the \textit{Milky Way-est} \citep{Buch2024}, the \textit{DREAMS} \citep{Rose2025} and the \textit{Auriga} simulation suites \citep{Grand2017, Grand2024}. As the rotation to the real MW--LMC orientation \citep[][appendix A]{Arora2024} can be easily generalised, this implies that we could apply our SBI framework to any cosmological MW--LMC analogue. This may be beneficial to start to build up statistics on e.g, the inferred reflex motion parameters of MW analogues for a variety of unique LMC analogue masses and infall scenarios. Or, this could even allow the existing sets of cosmological MW--LMC analogues to be used as the input simulations for the application of SBI to observational data.

\subsection{Application to observational data}\label{sec:discussion-applicationtodata}

This study has focused on presenting the framework to apply SBI to understand the dynamics of the MW in response to the recent infall of the LMC. We have demonstrated that we can use SBI to correctly recover known simulations parameters, e.g., the MW and LMC masses, across a range of rigid, deforming and cosmological MW--LMC models. Additionally, in Sec.~\ref{sec:posterior-diagnostic-checks}, through a selection of diagnostic tests, we showed that the SBI posteriors are reliable and unbiased. 

For the dynamical quantities considered in this work, there already exists observational measurements. For example, there is the DESI BHB \citep{Bystrom2025} and also the all-sky red giant stars out to $160 \, \rm kpc$ from the combined \textit{H3 + SEGUE + MagE} dataset \citep{Chandra2025}. 
Indeed, they measure the MW reflex motion parameters $v_{\rm travel},\, l_{\rm apex}$ and $b_{\rm apex}$, but they also report on the measurement of the Galactocentric solar motion corrected stellar radial velocities, $v_{\rm GSR}$. In both sets of observations, stars in the southern Galactic hemisphere exhibit negative mean radial velocities, i.e., $\langle v_{\rm GSR} \rangle < 0 \,\rm km\, \rm s^{-1}$, due to the perturbation from the LMC.
Interestingly, the same measurement for stars in the northern hemisphere disagrees between the studies. The DESI sample in \citet{Bystrom2025} shows that radial velocities remain consistent with the equilibrium expectation of $\langle v_{\rm GSR} \rangle \sim 0\, \rm km \, \rm s^{-1}$, whereas the \textit{H3 + SEGUE + MagE} sample in \citep{Chandra2025} displays a positive mean radial velocity i.e., $\langle v_{\rm GSR} \rangle > 0 \,\rm km\, \rm s^{-1}$. 
This is an intriguing discrepancy between the two surveys, and it is one which we have addressed with the application of our SBI framework to these datasets \citep[see,][]{Brooks2025}.
In our Figs.~\ref{fig:1} \& \ref{fig3} \& \ref{fig4} we show the on-sky $\langle v_{\rm GSR} \rangle$ distributions for the corresponding MW--LMC simulation stellar halo using stars beyond $50\,\rm kpc$ and subtracting off the mean motion across the entire sky. Generally, the strength of the average radial velocity signal will depend on the model for the MW and LMC. Plus, the boundaries of any survey's sky coverage could influence the results of averaging in each hemisphere e.g., the northern Galactic hemisphere in all of Figs.~\ref{fig:1} \& \ref{fig3} \& \ref{fig4} contains regions of both positive and negative average radial velocity. The existing discrepancies for the observational results for the reflex motion and the solar motion corrected stellar radial velocities \citep{Bystrom2025, Yaaqib2024, Chandra2025} motivate us to apply our SBI framework to this problem in order to better understand these dynamical quantities \citep[][]{Brooks2025}.

\section{Conclusions}\label{sec:conclusions}

In this work, we have presented the first use of an SBI framework to investigate the dynamics of the MW--LMC interaction. We vary model parameters including the MW mass, the LMC mass and the dynamical friction strength. From each simulation, we measure the observables of the induced MW reflex motion in response to the infalling LMC.
We have performed a series of diagnostic and validation checks across rigid, deforming, and cosmological MW--LMC models. We summarise our findings as follows:

\begin{enumerate}
    \item We have produced, and made publicly available via \href{https://zenodo.org/records/15356067}{Zenodo}, a set of $128,000$ unqiue MW--LMC simulations with stellar haloes evolved to present-day. These simulations act as the base to train our \textit{amortised} SBI framework to return posterior distributions on e.g., the LMC mass.
    \item We have validated our SBI framework on rigid, deforming and cosmological MW--LMC models, showing that we are able to correctly recover e.g., the infall LMC mass when information about the induced MW reflex motion is known.
    \item Our SBI framework trained on many low fidelity MW--LMC simulations retains enough of the relevant physics of the higher fidelity simulations to avoid model misspecification. This is promising for the application to real datasets as it allows for the rapid exploration of large model parameter spaces at a fraction of the computational cost required to run the higher fidelity simulations.
    \item We find that for the cosmological MW--LMC analogue, the dynamical friction strength is well constrained to values that are $\sim 2- 3$ times greater than, and intriguingly not consistent with, those expected by classical Chandrasekhar dynamical friction. For the deforming models, we cannot make similar statements as the dynamical friction strength is poorly constrained.
\end{enumerate}

\noindent
Overall, we have successfully validated our SBI framework to understand the MW dynamics across rigid, deforming, and cosmological MW--LMC models.
This motivates us to apply our methodology to real MW stellar halo data from the DESI \citep{Bystrom2025} and H3 + SEGUE + MagE datasets \citep{Chandra2025} to better understand properties of the MW, LMC, and the induced MW reflex motion \citep[][]{Brooks2025}. 

\section*{Acknowledgements}

RANB acknowledges support from the Royal Society and the Flatiron Institute, Simons Foundation. RANB would like to thank Jason Ran, Paul Shah, Zixiao Hu, Lorne Whiteway and Axel Widmark for useful discussions on simulation based inference. 
RANB thanks Kathryn Johnston for valuable feedback that improved the quality of the paper.
RANB thanks Arpit Arora for sharing the substructure-free FIRE simulation data and his code to rotate the \texttt{m12b} MW--LMC cosmological analogue into the real MW--LMC orientation. 
JLS acknowledges the support of the Royal Society (URF\textbackslash R1\textbackslash191555; URF\textbackslash R\textbackslash 241030). 
NGC acknowledges support from the Heising-Simons Foundation grant \#2022-3927, through which NGC is supported by the Barbara Pichardo Future Faculty Fellowship.

\section*{Data Availability}\label{sec:data-availability}

The MW--LMC simulation parameters and the data for $128,000$ unique stellar haloes, each with $4500$ particles, is available on Zenodo \href{https://zenodo.org/records/15356067}{here}.

\textit{Software:} \texttt{sbi} \citep{tejero-cantero2020sbi}, \texttt{agama} \citep{2019MNRAS.482.1525V}, \texttt{gala} \citep{gala}, NumPy \citep{harris2020array}, Matplotlib \citep{Hunter:2007}, Seaborn \citep{Waskom2021}, \texttt{corner} \citep{corner}, Astropy \citep{astropy:2013, astropy:2018, astropy:2022}, SciPy \citep{2020SciPy-NMeth}.



\bibliographystyle{mnras}
\bibliography{biblio} 





\bsp	
\label{lastpage}
\end{document}